# Tunable Magneto-optical Kerr effect in two-dimensional non-collinear antiferromagnetic material HfFeCl$_6$

Di Zhou[†], Ning Ding [†], Haoshen Ye, Shuai Dong, and Shan-Shan Wang[*]

*Key Laboratory of Quantum Materials and Devices of Ministry of Education, School of Physics, Southeast University, Nanjing 211189, P. R. China.*

[*]Corresponding author. Email: wangss@seu.edu.cn

**ABSTRACT**

With the development of two-dimensional (2D) magnetic materials, magneto-optical Kerr effect (MOKE) is widely used to measure ferromagnetism in 2D systems. Although this effect is usually inactive in antiferromagnets (AFM), recent theoretical studies have demonstrated that the presence of MOKE relies on the symmetry of the system and antiferromagnets with noncollinear magnetic order can also induce a significant MOKE signal even without a net magnetization. However, this phenomenon is rarely studied in 2D systems due to a scarcity of appropriate materials hosting noncollinear AFM order. Here, based on first-principles calculations, we investigate the HfFeCl$_6$ monolayer with noncollinear Y-AFM ground states, which simultaneously breaks the time-reversal ($T$) and time-inversion ($TI$) symmetry, activating the MOKE even though with zero net magnetic moment. In addition, four different MOKE spectra can be obtained in the four permutation states of spin chirality and crystal chirality. The MOKE spectra are switchable when reversing both crystal and spin chirality. Our study provides a material platform to explore the MOKE effect and can potentially be used for electrical readout of AFM states.

Reducing power consumption and scale has become of utmost importance in spintronics. 2D antiferromagnetic (AFM) materials have great potential in these regards, which makes them outstanding material candidates in the future of the next spintronic devices.[1-7] Unlike ferromagnetic (FM) materials that can be easily probed and manipulated, AFM materials have vanishing net magnetization. It makes them robust to external fields and thus renders them invisible to be probed and manipulated.[6-9]

The magneto-optical Kerr effect (MOKE) characterizes the state change of light when it interacts with a magnetic material,[10] which has been used to probe magnetic structures.[11-16] It is usually active in FM materials in the presence of spin-orbit coupling (SOC) and exchange splitting.[17-20] With the rising of 2D materials, MOKE can be used to detect magnetic 2D systems.[21-26] Experimentally, CrI$_3$ monolayer was first reported the MOKE signal and its origin are discussed by theoretically.[21] After that, large MOKE effects were also observed in other 2D FM systems, including Fe$_3$GeTe$_2$,[23] and Cr$_2$Ge$_2$Te$_6$ monolayers.[24] In general, MOKE is not active in collinear AFM systems due to the zero net magnetization.

Despite the existence of collinear AFM systems, there exist many noncollinear AFM materials, in which the spin moments tilt at angles ranging from 0 to π. Actually, the presence of MOKE relies on the symmetries of a system.[27] In FM systems, the active MOKE occurs due to the breaking of time-reversal symmetry. In collinear AFM systems, the MOKE signal is suppressed by the combination of time-reversal and inversion symmetry $TI$. However, in noncollinear AFM systems, the *TI* symmetry is often broken by the noncollinear spin structure.[27-31] Therefore, the MOKE signals may occur in noncollinear AFM systems even without net magnetization. Noncollinear antiferromagnets bulk Mn$_3$*X* were predicted to present large MOKE signals,[27,28] which were verified experimentally in Mn$_3$Sn[30] and Mn$_3$Ge[31]. In 2D noncollinear AFM systems, the MOKE is little studied, and only a few materials are reported, such as W$_3$Cl$_8$.[32] Hence, it would be interesting to explore MOKE and find a route to tune it in 2D noncollinear AFM systems.

In this work, based on first-principles calculations, we reveal the structure and magnetic properties of the HfFeCl$_6$ monolayer. Due to the triangular frustration in the system, coplanar noncollinear Y-type AFM is confirmed as the ground state. A pronounced Kerr signal can be observed, the physical origin of which is that the Y-type AFM order breaks the *TI* symmetry, causing the non-zero off-diagonal term $\sigma_{xy}$ of the optical conductivity tensor (OCT). Additionally, we investigate the modulation of the MOKE spectrum by crystal chirality arising from the arrangement of non-magnetic atoms and spin chirality arising from noncollinear AFM order. The sign of the MOKE spectrum is reversed when both crystal chirality and spin chirality are simultaneously reversed by mirror symmetry along *x*-axis ($M_x$). Furthermore, by rotating the Y-type spin configuration, a 120° periodic variation of the maximum value of the Kerr angle $\theta_K$ can be achieved. Our study here provides a platform to investigate the microscopic origin of the Kerr effect and it can be exploited in AFM spintronics.

The bulk phase of HfFeCl$_6$ was found in 1993,[33] which has a structure of alternating double layers of chlorine and metal atoms with space group of $P\bar{3}1c$ (No.163), shown in Fig. 1(a). The HfFeCl$_6$ layers are stacked along the [001] direction and the interlayer coupling is van der Waals (vdW) interaction. The cleavage energy is determined by calculating the total energy of the system as a function of the distance d for a monolayer. Figure 1(b) illustrates the results with the cleavage energy of 11.1 meV/Å$^2$, which is smaller than those of familiar 2D materials such as graphene (~ 26.8 meV/Å$^2$),[34] CrI$_3$ (~ 18.7 meV/Å$^2$),[35] implying the monolayer structure can be exfoliated from the bulk easily.

The top and side views of the HfFeCl$_6$ monolayer are presented in Fig. 1(c). The monolayer has a honeycomb structure similar to CrI$_3$,[21,35] but the magnetic atomic sites are occupied by two different transitional metals: iron (Fe) and hafnium (Hf), respectively, which form two nested triangular sublattices. The Fe and Hf layers are sandwiched between two Cl layers. Every Fe$^{2+}$ (also Hf$^{4+}$) ion and the surrounding six Cl$^-$ ions form an octahedral structure connected by edge-sharing mode. The optimized HfFeCl$_6$ monolayer has a lattice constant of $a = b = 6.39$ Å, and the lattice angles are $\alpha = \beta = 90°$, $\gamma = 120°$, with a non-polar space group $P312$ (No.149). The phonon spectrum was evaluated in a 2 × 2 supercell to verify its dynamic stability. The HfFeCl$_6$ monolayer is dynamically stable since no imaginary frequencies in the entire Brillouin zone, as shown in Fig. 1(d).

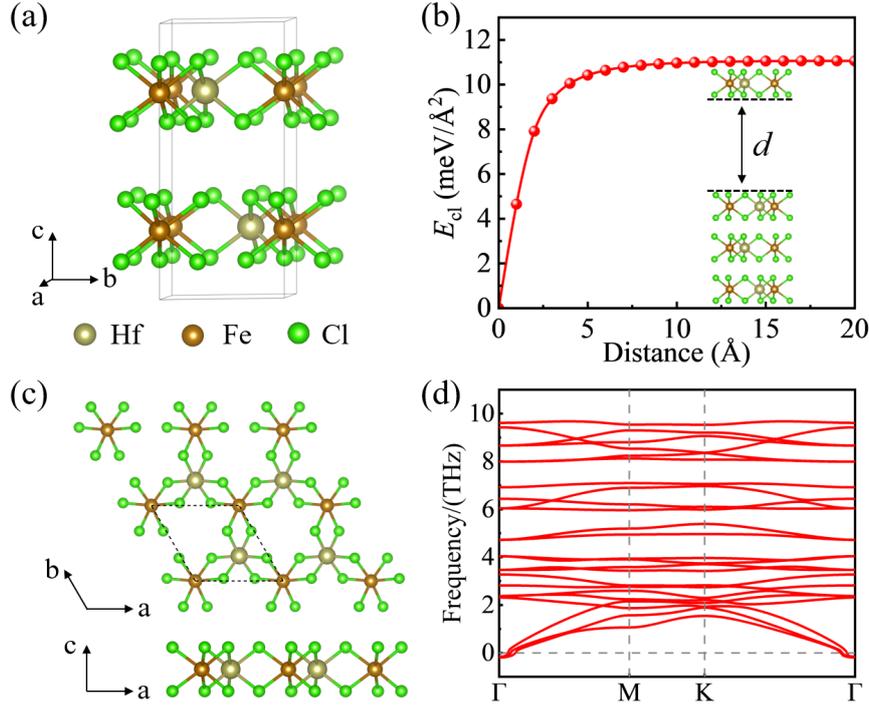

Fig. 1. (a) Bulk phase of HfFeCl$_6$. (b) Schematic of exfoliation process. (c) The top and side views of the HfFeCl$_6$ monolayer, showing the triangular sublattice and octahedral structure of Hf and Fe atoms. (d) Phonon spectrum of monolayer HfFeCl$_6$.

The nominal valences of Hf, Fe, and Cl are +4, +2, and −1, respectively. The 5$d$ orbitals of Hf$^{4+}$ are completely empty, while Fe$^{2+}$ has an electronic state of 3$d^6$, as confirmed by DFT calculations. The orbital-projected density of states (DOS) is presented in Fig. 2(a), showing that the five spin-up electrons fully occupy the five 3$d$ orbitals and one more spin-down electron occupies the $d_{z2}$ orbital. The highest occupied state is the spin-down $d_{z2}$ orbital. Thus, the magnetic properties of the HfFeCl$_6$ monolayer can be attributed to Fe$^{2+}$, which exhibits a localized magnetic moment with 4.00 $\mu$B in each Fe$^{2+}$. Four possible magnetic orders are calculated to determine the magnetic ground state, see Fig. S1 and Table S1 in Supplemental Material (SM). Considering the heavy element Hf, these magnetic configurations are also calculated with the consideration of SOC to double-check the ground state, as shown in Table S1. The results indicate that the Y-type AFM state with 120° angles considering the triangular sublattice of Fe$^{2+}$ is the most energetically favorable, as depicted in Fig. 2(b).

Magnetic anisotropy energy (MAE) is calculated. The energy for spin in the $xy$ plane is 0.43 meV lower than that of spin in the $xz$ and $yz$ plane in the formula unit, implying a magnetic anisotropy of easy-plane type.

Then, the exchange interactions between nearest-neighbor and next-nearest-neighbor Fe atoms (depicted in Fig.2(b)) have been calculated (see in Fig. S2 and Table S2 of SM). By comparing the DFT energies of different magnetic orders at normalized spins, the exchange interactions $J_1$ and $J_2$ can be derived as -0.20 meV and 0.09 meV, respectively. Such weak magnetic exchange effect is the result of the long distance between neighboring Fe pairs. A classical model Hamiltonian can be constructed as follows:

$$H = -J_1 \sum_{\langle i,j \rangle} \mathbf{S}_i \cdot \mathbf{S}_j - J_2 \sum_{\langle\langle i,m \rangle\rangle} \mathbf{S}_i \cdot \mathbf{S}_m - A \sum_i (\mathbf{S}_i^z)^2 \qquad (1)$$

where $\mathbf{S}_i$ represents the normalized spin ($|\mathbf{S}| = 1$), and $A$ represents the anisotropy constant. Based on the above DFT coefficients and Eq. (1), Monte Carlo (MC) simulation of the monolayer has been performed to verify the magnetic ground state, which is clearly described in Section IV of SM. The MC snapshot at a temperature of 0.01 K (Fig. 2(c)) confirms that the noncollinear Y-AFM magnetic configuration is indeed the magnetic ground state. In addition, the peak at $T_N \sim$ 2.4 K of the heat capacity indicates the magnetic phase transition as depicted in Fig. 2(d).

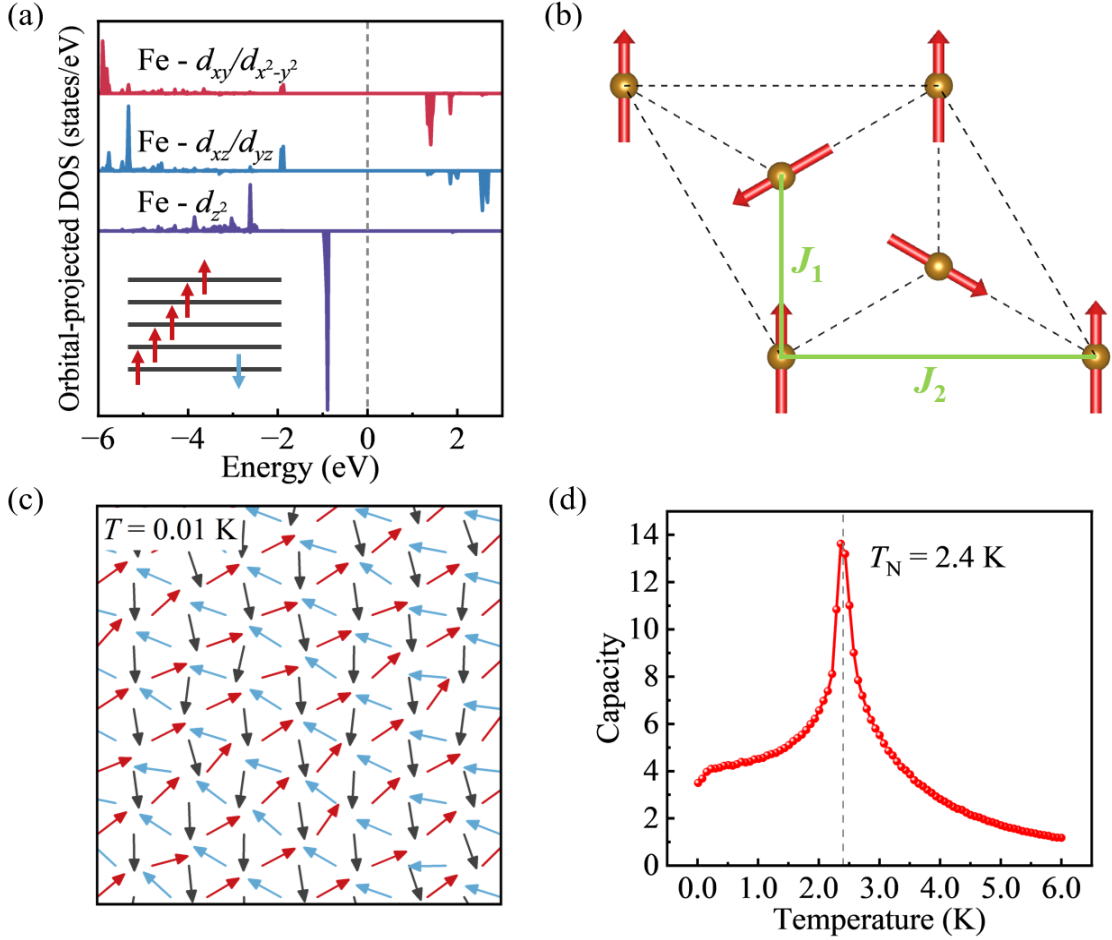

Fig. 2. (a) The orbital-projected density of states (DOS) of HfFeCl$_6$ monolayer. (b) The schematic diagram of Y-type AFM. The arrow shows the magnetic moment orientation of each Fe ion. $J_1$ and $J_2$ denote the exchange couplings between the nearest-neighboring and next-nearest-neighboring Fe$^{2+}$. (c) The Monte Carlo snapshot of the HfFeCl$_6$ monolayer. (d) The MC simulations for magnetic heat capacity $C$ and magnetic transitions.

Magneto-optical Kerr effect is activated when both time-reversal ($T$) symmetry and time-reversal combined with space-reversal ($TI$) symmetry are broken. The Y-type AFM state in HfFeCl$_6$ monolayer naturally breaks both symmetries, allowing for the emergence of the MOKE signal. The easy magnetization plane of the HfFeCl$_6$ monolayer is the layer plane. According to the angular relationship between the magnetization intensity, the surface of the medium, and the plane of incidence, the HfFeCl$_6$ monolayer has a polar MOKE. The complex Kerr angle of the monolayer with threefold rotational symmetry is given as:[36]

$$\phi_K = \theta_K + i\eta_K = \frac{2\xi_{xy}}{1 - (\xi_{xx} + n_s)^2 - \xi_{xy}^2} \quad (2)$$

where the real part $\theta_K$ is the Kerr rotation and the imaginary part $\eta_K$ is the Kerr ellipticity. The dimensionless optical conductivity is $\xi_{xy} = \sigma_{xy}Z_0$ and $\xi_{xx} = \sigma_{xx}Z_0$, where $Z_0$ is the vacuum impedance, and $n_s$ is the substrate refractive index. The elements $\sigma_{xy}$ and $\sigma_{xx}$ in the optical conduction tensor (OCT) can be obtained via first principles,[36] where the off-diagonal term $\sigma_{xy}$ reflects the presence or absence of the Kerr signal.

The magnetic point group (MPG) is 32', which has a threefold rotational symmetry, so that the optical conductivity tensor (OCT) $\boldsymbol{\sigma}$ can be expressed as:[37]

$$\boldsymbol{\sigma}\,(32') = \begin{pmatrix} \sigma_{xx} & \sigma_{xy} & 0 \\ -\sigma_{xy} & \sigma_{xx} & 0 \\ 0 & 0 & \sigma_{zz} \end{pmatrix} \quad (3)$$

According to Eq. (2), the off-diagonal term $\sigma_{xy} \neq 0$, indicates the presence of the MOKE signal. Our calculation verifies a magneto-optical Kerr angle in the monolayer, which reaches a maximum at an incident light of 3.8 eV, as shown in Fig. 3(a). When the time-reversal operation $T$ is performed, the MOKE spectrum is completely opposite, as presented in Fig. 3(b). This opposition derives from the transformation rules of the optical conductivity tensor $\boldsymbol{\sigma}$. According to Onsager's relation,[38] the $\boldsymbol{\sigma}$ tensor transforms under $T$ symmetry operation as follows:

$$\boldsymbol{\sigma} \xrightarrow{T} T\boldsymbol{\sigma}T^{-1} = \boldsymbol{\sigma}^{\mathrm{t}} \quad (4)$$

where "t" donates the transpose of $\boldsymbol{\sigma}$. The off-diagonal term $\sigma_{xy}$ changes the sign, thus implying that the Kerr angle is also inversed. Under $I$ symmetry, the $\boldsymbol{\sigma}$ tensor transforms as follows:

$$\boldsymbol{\sigma} \xrightarrow{I} I\boldsymbol{\sigma}I^{-1} = \boldsymbol{\sigma} \quad (5)$$

where $\boldsymbol{\sigma}$ is invariant under inversion operation $I$.

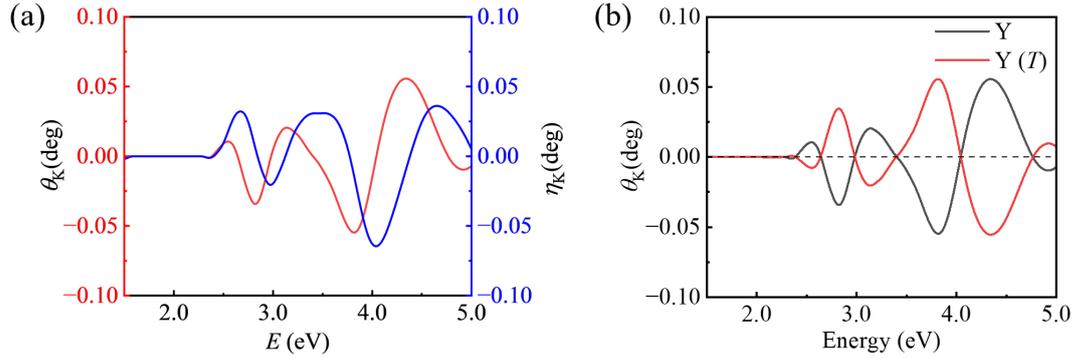

Fig. 3. (a) Kerr angle $\theta_K$ and ellipticity $\eta_K$ of HfFeCl$_6$ monolayer at OCT of 32'. (b) Kerr angle $\theta_K$ under $T$ symmetry operation.

The chiral space group of $P312$ and the magnetic point group of 32' in the HfFeCl$_6$ monolayer suggest that there are two types of chirality in the monolayer, namely crystal chirality, and spin chirality.

The crystal chirality comes from different arrangements of non-magnetic atoms. In the case of the HfFeCl$_6$ monolayer, the space group of $P312$ is non-centrosymmetric and thus it has a global crystal chirality.[39,40] The top- and side-view schematics of the lattice structure under the crystal chirality $\chi = +1$ and $\chi = -1$, are presented in Fig. 4(a), and the difference originates from the positional changes of the non-magnetic atoms Hf and Cl.

While spin chirality stems from coplanar noncollinear AFM of the triangular sublattice constituted by magnetic atoms, it can be characterized by vector spin chirality as:[41]

$$\kappa = \frac{2}{3\sqrt{3}} \sum_{\langle ij \rangle} [S_i \times S_j]_z \tag{6}$$

where $\langle ij \rangle$ runs over the nearest-neighboring Fe spins. As indicated in Fig. 4(b), the sign of $\kappa$ represents the two vector spin chiral states in a counterclockwise sequence, i.e., the right-handed (anticlockwise) state for $\kappa = +1$ and the left-handed (clockwise) state for $\kappa = -1$. For both right- and left-handed chirality, different spin configurations can be produced by simultaneously rotating the spin in the plane, as depicted in Fig. 4(b), which shows a schematic of the case of rotation angle $\varphi = 30°$. The vector spin chirality and spin rotation discussed here allow us to characterize various antiferromagnetic configurations with 120° coplanar noncollinear magnetic order.

Here we focus on the MOKE of four chiral states which can be described by permutations of crystal and spin chirality, namely ($\chi = +1$, $\kappa = +1$), ($\chi = +1$, $\kappa = -1$), ($\chi = -1$, $\kappa = +1$), and ($\chi = -1$, $\kappa = -1$). All four states have the same MPG of 32'. The symmetry operations among these four states are summarized in Fig. 4(c), where the $C$ operation is purely a chirality switch of $\chi$ or $\kappa$. Two states with exactly opposite signs of both chiral $\chi$ and $\kappa$ can be connected by mirror symmetry operations along x-axis ($M_x$). For example, when the $M_x$ operation is performed on ($\chi = +1$, $\kappa = +1$) state, it turns to ($\chi = -1$, $\kappa = -1$) state, meanwhile, the Kerr spectrum is completely reversed, as displayed in Fig. 4(d). The same reversal also occurs in the transition from ($\chi = -1$, $\kappa = +1$) state to ($\chi = +1$, $\kappa = -1$) state. This reversal stems from the fact that the $M_x$ operation can be equated to the $C_{2z}M_zT$ operation, namely $TI$ operation, where the $C_{2z}$ and $M_z$ operations do not change the sign of the off-diagonal term $\sigma_{xy}$ in the optical conductivity tensor. Thus, it has the same MOKE spectrum under $T$ and $M_x$ operations, which means that $M_xT$ operation doesn't change the Kerr angle.

The single reversal of a particular chirality leads to a change in the profile of the spectrum. At a given $\kappa = +1$, both $\chi = +1$ and $\chi = -1$ lead to the MPG 32', but their MOKE signals are not identical, as displayed by the black solid line and the red solid line in Fig. 4(d). The maximum value of $\theta_K$ is 0.057° (at 4.34 eV) for $\chi = +1$, while the maximum is 0.069° (at 4.59 eV) for $\chi = -1$, which is 21% larger than the former.

With the same chiral combinatorial state, the MOKE signal can be affected by the azimuthal angle $\varphi$ of the Y-type magnetic configuration. Since the chirality switch doesn't change the MPG, we choose the ($\chi = +1$, $\kappa = +1$) state as a reference and focus on the effect of $\varphi$ on the MPG and MOKE spectrum, as listed in Table I. In the interval $\varphi$ from 0° to 120°, new MPGs of 32.1 and 3 appear. Their OCTs can be expressed as:[37]

$$\sigma(3) = \begin{pmatrix} \sigma_{xx} & \sigma_{xy} & 0 \\ -\sigma_{xy} & \sigma_{xx} & 0 \\ 0 & 0 & \sigma_{zz} \end{pmatrix} \tag{7}$$

$$\sigma(32.1) = \begin{pmatrix} \sigma_{xx} & 0 & 0 \\ 0 & \sigma_{xx} & 0 \\ 0 & 0 & \sigma_{zz} \end{pmatrix} \tag{8}$$

According to Eq. (2), the emergence of the MOKE signal relies on non-zero off-diagonal terms $\sigma_{xy}$.

When $\varphi$ equals an odd multiple of 15°, a non-zero $\sigma_{xy}$ in MPG 3 indicates the presence of MOKE signal. When $\varphi = 30°$ and 90°, a zero $\sigma_{xy}$ in MPG 32.1 indicates the absence of MOKE signal. In this case, there is a spin perpendicular to the $M_x$ mirror plane among three spins with different orientations, of which direction is invariant under the $M_x$ operation, as shown in Fig. 5(a). The Kramers theorem is maintained by the combined $TM_x$ symmetry, and therefore the MOKE signal vanishes. The MOKE spectra and maximum value of $\theta_K$ of different $\varphi$ are displayed in Fig. 5(b) and (c), respectively. It can be seen that $\theta_{Kmax}$ as a function of $\varphi$ has a periodicity of 120°. In addition, for two different azimuthal angles $\varphi_1$ and $\varphi_2$, the MOKE spectra are the same if $\varphi_1 + \varphi_2 = 120°$, and opposite if $\varphi_1 - \varphi_2 = 60°$. The former originates from the threefold rotational symmetry of the system, while the latter is the result of $Tt$[1/3, -1/3, 0] symmetry operation between $\varphi_1$ and $\varphi_2$, where $t$ is translational symmetry. This provides an effective alternative way to achieve MOKE spectral inversion beyond time-reversal operations $T$ and chiral switch.

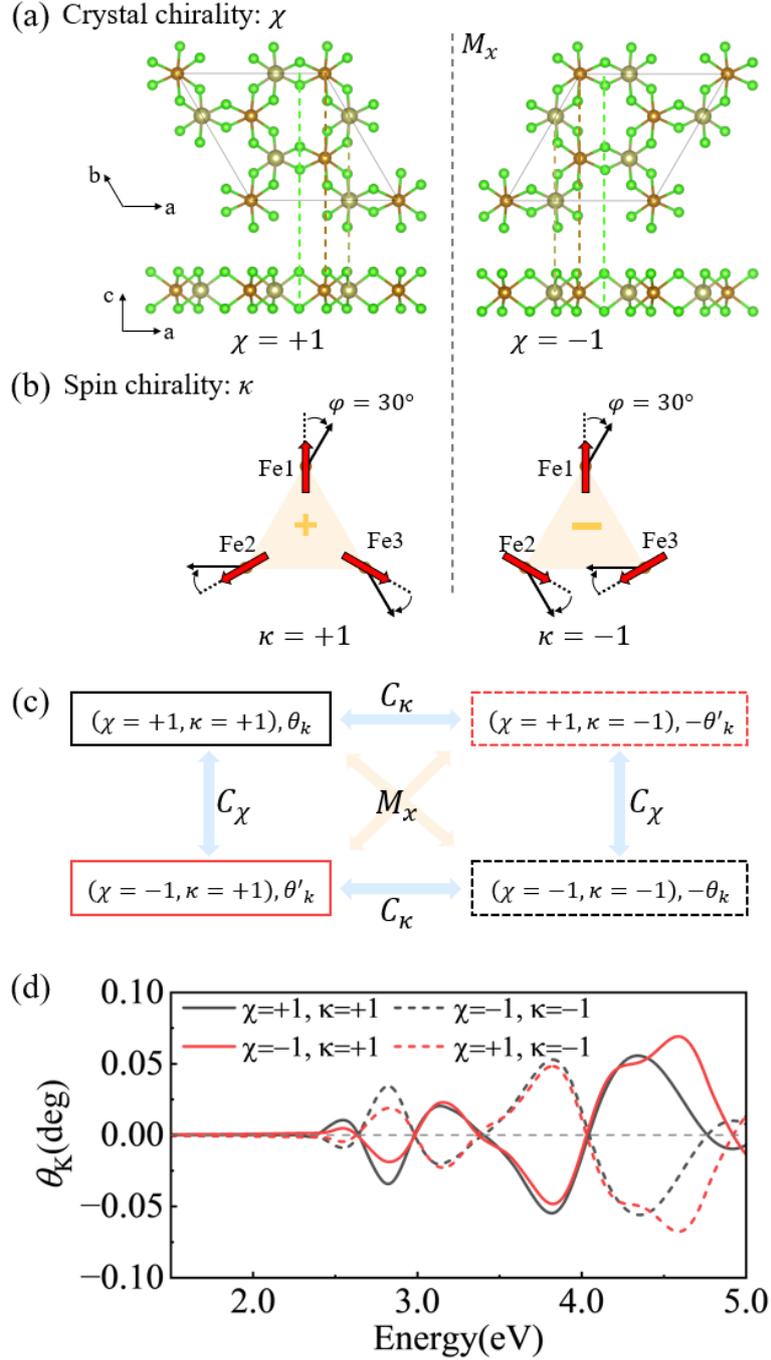

Fig. 4. (a) The top- and side-view of the crystal structures of the $HfFeCl_6$ monolayer with the crystal chiralities $\chi = +1$ and $\chi = -1$. The dotted lines mark the relative positions of the Hf, Fe, and Cl atoms. (b) Spin configurations on the Fe triangular sublattice with the right-handed ($\kappa = +1$) and left-handed ($\kappa = -1$) spin chirality. The red arrows indicate the magnetic moment orientation of each Fe atom. $\kappa$ is calculated for the downward triangle in the counterclockwise sequence. The azimuthal angle $\varphi$ is defined to measure the clockwise synchronous rotation of all spins, with the black solid line marking the position at $\varphi = 30°$. (c) The symmetry relations under different operations namely $M_x$ (mirror symmetry operations along x-axis), $C_\chi$ and $C_\kappa$ (the pure chirality switch of $\chi$ or $\kappa$) between the four states ($\chi = +1, \kappa = +1$), ($\chi = +1, \kappa = -1$), ($\chi = -1, \kappa = +1$), and ($\chi = -1, \kappa = -1$). Here, $\theta_K$ and $\theta'_K$ represent different Kerr spectra. (d) the Kerr spectrum under the four chirality states.

Table I. Magnetic point groups (MPG) and off-diagonal term $\sigma_{xy}$ for HfFeCl$_6$ monolayers in the ($\chi = +1$, $\kappa = +1$) state for different magnetic states with azimuthal angle $\varphi$ in the range from 0° to 120°. The symbol '√' represents a non-zero $\sigma_{xy}$ term and an observable MOKE, while '×' represents a zero $\sigma_{xy}$ term and an unobservable MOKE.

| $\varphi/°$ | 0 | 15 | 30 | 45 | 60 | 75 | 90 | 105 | 120 |
|---|---|---|---|---|---|---|---|---|---|
| **MPG** | 32' | 3 | 32.1 | 3 | 32' | 3 | 32.1 | 3 | 32' |
| $\sigma_{xy}$ | √ | √ | × | √ | √ | √ | × | √ | √ |

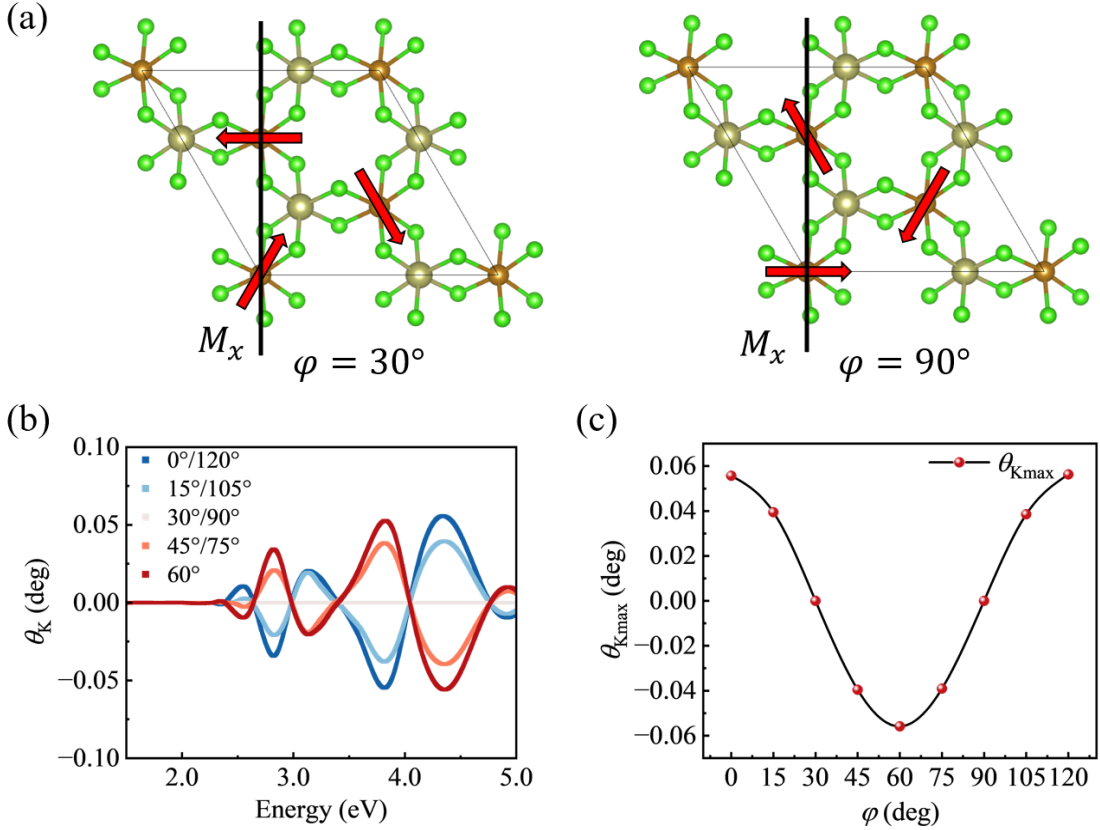

Fig. 5. (a) Schematic of magnetic configurations and $M_x$ plane in the monolayer at $\varphi = 30°$ and $\varphi = 90°$. The MOKE spectra (b) and maximum value of $\theta_K$ (c) of different $\varphi$ at the range from 0° to 120°.

In conclusion, based on first-principles calculations, we investigate HfFeCl$_6$ monolayer with a nested triangular lattice. Our results demonstrate that HfFeCl$_6$ monolayer keeps dynamically stable and the ground state favors the coplanar noncollinear Y-type AFM state within the $xy$ easy plane, which is further verified by MC simulations. Such noncollinear AFM magnetic configuration naturally breaks the $T$ and $TI$ symmetry, allowing the emergence of MOKE signal even though without net magnetic moment. The sign of the off-diagonal term $\sigma_{xy}$ of OCT can be reversed under the $T$ operation, which determines the sign of the MOKE spectrum. Interestingly, such inversion can also occur when both crystal chirality and spin chirality are simultaneously reversed. The physical origin is that the $M_x$ operation in this system is equivalent to the $C_{2z}M_zT$ operation. When only the crystal chirality or spin chirality is changed, the shape of the MOKE spectrum changes slightly, reflecting the fact that both magnetic and non-magnetic atoms affect MOKE. Moreover, the system with different rotation angles of spin configuration has different MPG and OCT, resulting in changes of Kerr angles, which have a period of 120°. This study not only proves the existence of MOKE in the 2D system with noncollinear AFM, but also provids hints to future devices based on magneto-optical materials.

See the supplementary material that provides the computational methods for the first principles calculation and Mento Carlo simulation, and calculations of the magnetic configurations and exchange interactions.

ACKNOWLEDGMENTS

This work supported by the National Natural Science Foundation of China (Grant No. 12104089), Jiangsu Funding Program for Excellent Postdoctoral Talent under Grant No. 2024ZB001, and China Postdoctoral Science Foundation under Grant Number 2024M760423. We thank the Big Data Center of Southeast University for providing the facility support on the numerical calculations.

AUTHOR DECLARATIONS

Di Zhou and Ning Ding contributed equally to this work.

Conflict interest

The authors have no conflicts to disclose.

DATA AVAILABILITY

The data that support the findings of this study are available within the article and its supplementary material.